\documentclass[aps,twocolumn,superscriptaddress,showpacs,showkeys,preprintnumbers,amsmath,amssymb]{revtex4}

\usepackage{txfonts}
\usepackage{dcolumn}
\usepackage{mathrsfs}
\usepackage{bm}
\usepackage{amsmath,amssymb,epsfig,float,graphics}

\begin{document}

\title{Inhibiting decoherence of two-level atom in thermal bath by presence of boundaries}
\author{Xiaobao Liu}
\affiliation{Department of Physics, Key Laboratory of Low Dimensional Quantum Structures and Quantum Control of Ministry of Education, and Synergetic Innovation Center for Quantum Effects and Applications, Hunan Normal University, Changsha, Hunan 410081, P. R. China.}

\author{Zehua Tian}
\affiliation{Institute of Theoretical Physics, University of Warsaw, Pasteura 5, 02-093 Warsaw, Poland}

\author{Jieci Wang}
\affiliation{Department of Physics, Key Laboratory of Low Dimensional Quantum Structures and Quantum Control of Ministry of Education, and Synergetic Innovation Center for Quantum Effects and Applications, Hunan Normal University, Changsha, Hunan 410081, P. R. China.}

\author{Jiliang Jing\footnote{Corresponding author, Email: jljing@hunn.edu.cn}}
\affiliation{Department of Physics, Key Laboratory of Low Dimensional Quantum Structures and Quantum Control of Ministry of Education, and Synergetic Innovation Center for Quantum Effects and Applications, Hunan Normal University, Changsha, Hunan 410081, P. R. China.}

\begin{abstract}
We study,  in the paradigm of open quantum systems, the dynamics of quantum coherence of a static polarizable two-level atom which is coupled with a thermal bath of fluctuating electromagnetic field in the absence and presence of boundaries. The purpose is to find the conditions under which the decoherence can be inhibited effectively. We find that without boundaries, quantum coherence of the two-level atom inevitably decreases due to the effect of thermal bath. However, the quantum decoherence, in the presence of a boundary, could be effectively inhibited when the atom is transversely polarizable and near this boundary.
In particular, we find that in the case of two parallel reflecting boundaries, the atom with a parallel dipole polarization at arbitrary location between these two boundaries will be never subjected to decoherence provided we take some special distances for the two boundaries.
\end{abstract}

\pacs{\emph{03.65.Yz, 11.10.Wx, 03.65.Ta, 03.67.Hk}}
\keywords{Inhibit decoherence; Thermal bath; Reflecting boundary; Open quantum systems.}
\maketitle

\section{Introduction}
Quantum coherence is one of fundamental concepts in physics and a key quantum features of quantum mechanics beyond classical physics~\cite{Leggett}. The formulation of the notion of coherence trace back to the quantum optics~\cite{Glauber,Sudarshan}, which provide a number of methods to understand coherence.
The development of quantum coherence has provided a significant physical resource in quantum information~\cite{Nielsen} and in biological systems~\cite{Huelga},
for interpretation of the quantum reference frames and the asymmetry~\cite{Marvian,Bartlett}, and for applications in thermodynamics~\cite{Lostagio,Lostaglio1} and quantum metrology~\cite{Giovannetti,Demkowicz}. Moreover, {\AA}berg reported that the quantum coherence can be turn into a catalyst resource~\cite{berg}, which has important consequences in quantum thermodynamics. Most recently, a number of computable measures have been put forward to quantifying coherence,  such as the \textit{$l_1$} norm and the relative entropy of coherence\cite{Baumgratz}, and others suggested  the skew information~\cite{Girolami}. Mathematically, the coherence of a state is defined as its off-diagonal elements. For instance,  the \textit{$l_1$} norm of coherence and  relative entropy of coherence  for any state $\hat{\rho}$ are defined as
\begin{equation}\label{norm0}
C_{l_1}(\hat{\rho})=\sum_{\substack{i,j\\ i\ne j}}|\rho_{i,j}|\;,
\end{equation}
and
\begin{equation}\label{entropy0}
C_{\rm RE}(\hat{\rho})=S(\hat{\rho}_{\rm diag})-S(\hat{\rho})\;,
\end{equation}
respectively. In Eq.~(\ref{entropy0}),  $S(\hat{\rho})=-\textbf{Tr}(\hat{\rho}\log\hat{\rho})$  donates the von Neumann entropy, `` log " indicates logarithms which are taken to base two, and $\hat{\rho}_{\rm diag}$ is the matrix only containing the diagonal elements of $\hat{\rho}$.

At this point, quantum coherence is the crucial resource for quantum information technology and has significant meaning to the physical implementation.
From the point of view of practice, it is worth noting that the quantum resource will be subjected to destroyed unavoidably when a system coupled with the surrounding environment during the whole evolution~\cite{Breuer,Schlosshauer,Wang1,Tian,Liu,Wang2,Tian1,Tian2,Jia}. Therefore,
how to protect this important resource of quantum coherence from the external noise is an essential problem. So far, a lot of researches have been devoted to the protection of quantum coherence.
For example, it has been analyzed under dynamical conditions the coherence and quantum correlations of an open quantum system are totally unaffected  under the bit flip, bit-phase flip and phase flip channels \cite{thomas,Cianciaruso}, whose existence has been recently experimentally verified \cite{Silva}.
It has been recently showed that suitable non-Markovian structured environments can efficiently preserve quantum coherence and entanglement \cite{Man,Lo,Man1,Brito,Rodriguez,Gonzalez}.
In addition, how to protect quantum coherence of a two-level atom from the vacuum fluctuation of quantum fields has been discussed in Ref. \cite{Liu}.

It is noteworthy that the atom-field system has been widely adapted to achieve the quantum information processing, such as the quantum communication~\cite{Guan,Nielsen} and the preparation of quantum state ~\cite{Remote1,Remote2}. The quantum coherence is the key resource in quantum information processing, so it becomes very important for us to protect the coherence effectively. In this paper, we will investigate the dynamics of quantum coherence of a static polarizable two-level atom in a thermal bath of electromagnetic field at finite temperature $T$.  Our aim is to find out how to protect quantum coherence effectively from the effect of thermal bath. We demonstrate that without boundaries, the two-level atom inevitably suffers from decoherence due to the interaction with the thermal bath and the decoherence will be enhanced with the increases of temperature $T$, meaning that the quantum coherence will be damaged more. However, by introducing reflecting boundaries we show that the decoherence can be inhibited effectively under certain conditions.

The organization of the paper is as follows. In Sec. II we give a review of the general formalism of quantum master equation which a two-level atom coupled with quantized electromagnetic field in a thermal bath of temperature $T$  and the atomic state is a maximally coherent state initially. In Sec. III, we calculate the quantum coherence in the case of without boundaries. In Sec. IV, we find that the decoherence can be inhibited effectively under appropriate condition by the presence of a reflecting boundary. Furthermore, in Sec. V we study the quantum coherence in the presence of two reflecting plates and analyze how to protect it totally under certain conditions.
Finally, we will give a summary of our conclusions in Sec. VI.

\section{the quantum master equation}

Let us start with the evolution of a static two-level atom which is coupling with quantized electromagnetic field. We assume that the total Hamiltonian of the atom-field system takes the form~\cite{Compagno}
\begin{eqnarray}
H=H_{S}+H_{F}+H_{I}\;,
\end{eqnarray}
where $H_{S}=\frac{1}{2}\hbar\omega_{0}\sigma_{3}$ denotes the Hamiltonian of the two-level atom, $H_{F}$ is the Hamiltonian of the electromagnetic field, and $H_{I}(\tau)=-e\textbf{r}\cdot\textbf{E}(x(\tau))$ is the interaction between the atom and the field. Here,  $\omega_{0}$ represents the transition frequency of the two-level atom, $\sigma_{3}$ is the Pauli matrix, $e\textbf{r}$ is the electric dipole moment of the atom, and $\textbf{E}(x)$ denotes the electric field strength.

Initially, the state of the atom-field system is assumed to be $\hat{\rho}_{tot}(0)=\rho(0)\otimes|0\rangle\langle0|$, where $\hat{\rho}(0)$ is the initial reduced density matrix of the atom and $|0\rangle\langle0|$ denotes the vacuum of electromagnetic field.
In the frame of the atom, the evolution of the total system obeys the von Neumann equation~\cite{Breuer}
\begin{eqnarray}\label{master0}
\frac{d}{d \tau}\hat{\rho}_{tot}(\tau)=-\frac{i}{\hbar}[H,\hat{\rho}_{tot}(\tau)]\;,
\end{eqnarray}
where $\tau$ is the proper time of the atom. Because our main interest here is the evolution of the atom and to obtain that we will trace over the field degrees of freedom, i.e., $\rho(\tau)={\rm Tr}_F[\rho_{tot}(\tau)]$. After some simplifications, the reduced state density of the atom, in the limit of weak coupling, is found to obey an equation in Kossakowski-Lindblad form~\cite{Gorini,Benatti}
\begin{equation}\label{master}
\frac{d}{d \tau}\hat{\rho}(\tau)=-\frac{i}{\hbar}[H_{eff},\hat{\rho}(t)]+\mathcal{L}[\hat{\rho}(\tau)]\;,
\end{equation}
where
\begin{eqnarray}
H_{eff}=\frac{1}{2}\hbar\Omega\sigma_{3}=\frac{\hbar}{2}\{\omega_{0}+\frac{i}{2}[\mathcal{K}(-\omega_{0})-\mathcal{K}(\omega_{0})]\}\sigma_{3}\;,\nonumber\\
\end{eqnarray}
is the effective Hamiltonian by absorbing the Lamb shift term, in which the $\Omega$ is the effective energy level-spacing of the atom, and
\begin{eqnarray}
\mathcal{L}[\hat{\rho}]=\frac{1}{2}\sum_{i,j=1}^{3}a_{ij}[2\sigma_{j}
\hat{\rho}\sigma_{i}-\sigma_{i}\sigma_{j}\hat{\rho}-\hat{\rho}\sigma_{i}\sigma_{j}]\;,
\end{eqnarray}
is the dissipator. Here,
\begin{equation}
a_{ij}=A\delta_{ij}-iB\epsilon_{ijk}\delta_{k3}-A\delta_{i3}\delta_{j3}\;,
\end{equation}
is the the coefficients of the Kossakowski matrix, in which
\begin{equation}
A=\frac{1}{4}[\mathcal{G}(\omega_{0})+\mathcal{G}(-\omega_{0})]\;,
\;\;B=\frac{1}{4}[\mathcal{G}(\omega_{0})-\mathcal{G}(-\omega_{0})]\;,
\end{equation}
where
\begin{eqnarray}\label{transforms}
&&\mathcal{G}(\lambda)=\int_{-\infty}^{\infty}d\triangle\tau e^{i\lambda\triangle\tau}G^+({\triangle\tau})\;,\nonumber \\
&&\mathcal{K}(\lambda)=\frac{P}{\pi i}\int_{-\infty}^{\infty}d\omega\frac{\mathcal{G}(\omega)}{\omega-\lambda}\;,
\end{eqnarray}
are the Fourier and Hilbert transforms to the electromagnetic field correlation function  $G^{+}(x-x')=\frac{e^{2}}{\hbar^{2}}\sum_{i,j=1}^{3}\langle+|r_{i}|-\rangle \langle-|r_{j}|+\rangle\langle0|\textbf{E}_{i}(x)\textbf{E}_{j}(x')|0\rangle$ respectively, in which $\lambda$ is the variable of the functions. Let us note that $r_i$ with $i\in\{1, 2, 3\}$ denotes the three components of the vector which points from the location of negative charge to that of positive charge.

If a two-level atom is initially prepared in a maximally coherent state, $|\psi(0)\rangle=\frac{1}{\sqrt{2}}(|+\rangle+|-\rangle)$, with  $|-\rangle$ and $|+\rangle$ being the ground state and excited state of the atom respectively, then according to Eq.~(\ref{master}), we can find that its time-dependent reduced density matrix is
given by
\begin{equation}\label{matrix}
\hat{\rho}=
\left(
\begin{array}{cc}
\frac{1}{2}+\frac{B}{2A}(e^{-4A\tau}-1)& \frac{1}{2}e^{-2A\tau-i\Omega\tau}\\
\frac{1}{2}e^{-2A\tau+i\Omega\tau}&\frac{1}{2}-\frac{B}{2A}(e^{-4A\tau}-1))
\end{array}
\right).
\end{equation}
Let us note that in Eq.~(\ref{matrix}), $(4A)^{-1}$ describes the time scale for atomic transition and $(2A)^{-1}$ is the time scale for the off-diagonal elements of the density-matrix ("coherence") decay ~\cite{Nielsen}.
We want to point out that the larger $(2A)^{-1}$ means that the quantum coherence will be decayed more slowly. In particularly,  when $(2A)^{-1} \rightarrow \infty$, the decoherence never happens and the coherence of the quantum state is constant.

\section{Decoherence of two-level atom in thermal bath  without  boundaries}

Let us now study how the electromagnetic field without  boundaries at finite temperature $T$ influence the quantum coherence. To do so, we will use the two-point function for the free electromagnetic field at finite temperature, which is given by~\cite{Brown,Yu},\begin{widetext}
\begin{eqnarray}\label{free}
\langle\textbf{E}_{i}(x(\tau))\textbf{E}_{j}(x(\tau'))\rangle_{f}=\frac{\hbar c}{4\pi^{2}\varepsilon_{0}}\sum_{n=-\infty}^{\infty}\frac{\delta_{ij}\partial_{0}\partial_{0}'-\partial_{i}\partial_{j}'}{(x-x')^{2}+(y-y')^{2}+(z-z')^{2}-(c t-c t'+i n\beta-i\varepsilon)^{2}}\;,
\end{eqnarray}
\end{widetext}
where $\varepsilon\rightarrow0^{+}$, $\partial_{j}'$ is the differentiation with respect to $x_{j}'$, $\varepsilon_{0}$ denotes the vacuum dielectric constant,  $\beta=\hbar c/Tk_{B}$ is the thermal wavelength, and the subscript `` f " represents the part induced by the unbounded space.

In the laboratory frame, for the two-level atom at rest, as a function of the proper time, the trajectory of the atom can be described by the equations
\begin{eqnarray}\label{trajectory}
t(\tau)=\tau\;,\;\;x(\tau)=x_{0}\;,\;\;y(\tau)=y_{0}\;,\;\;z(\tau)=z_{0}\;.
\end{eqnarray}
Applying Eqs.~(\ref{free}) and ~(\ref{trajectory}) to Eq.~(\ref{transforms}), we can obtain the Fourier transform of the correlation functions~\cite{Breuer}
\begin{eqnarray}\label{Fourier1}
\mathcal{G}_{f}(\lambda)=
\begin{cases}
\;\sum_{i=1}^{3}\frac{e^{2}|\langle-|r_{i}|+\rangle^{2}\lambda^{3}}{3\pi\varepsilon_{0}\hbar c^{3}}(N(\lambda)+1),\;\;(\lambda>0)\;,\\
\;\sum_{i=1}^{3}\frac{e^{2}|\langle-|r_{i}|+\rangle^{2}|\lambda|^{3}}{3\pi\varepsilon_{0}\hbar c^{3}}N(|\lambda|),\;\;(\lambda <0)\;.
\end{cases}
\end{eqnarray}
Thus, the coefficients of the Kossakowski matrix $a_{ij}$ can be expressed as
\begin{eqnarray}
A_{f}=\frac{\gamma_0}{4}[2N(\omega_{0})+1]\;,
\end{eqnarray}
and
\begin{eqnarray}
B_{f}=\frac{\gamma_0}{4}\;,
\end{eqnarray}
where $N(\omega_0)=\frac{1}{[\texttt{exp}(\omega_0\beta / c)-1]}$ represents the Planck distribution at the transition frequency $\omega_{0}$, $\gamma_0=\frac{e^2|\langle -|\textbf{r}|+\rangle|^2\,\omega_0^3}{3\pi\varepsilon_0\hbar c^3}$ describes the free-space spontaneous emission rate at zero temperature, and $\gamma_{f}=\gamma_0[2N(\omega_{0})+1]$ is the total transition rate. Hereafter, to simplify we let $N=N(\omega_0)$. It is worth noting that  $\tau_{f}=\frac{1}{2\gamma_0(2N+1)}$ is the time scale for the off-diagonal elements of the density-matrix  decay without boundaries.

By using the definition of the \textit{$l_1$} norm and relative entropy of coherence,  we can obtain the coherence of the atom interacting with a thermal bath of electromagnetic field at temperature $T$, which is given by
\begin{eqnarray}
C_{\textit{$l_1$}}(\hat{\rho})_{f}=\left|e^{-\frac{\tau}{\tau_{f}}} \right|=\left|(1-q)^{(N+\frac{1}{2})}\right|\;,
\end{eqnarray}
and
\begin{eqnarray}
\nonumber
C_{\rm RE}(\hat{\rho})_{f}&=&-M\log_{2}M-(1-M)\log_{2}(1-M)
\\
&&+\lambda_{+}\log_{2}\lambda_{+}
+\lambda_{-}\log_{2}\lambda_{-}\;,
\end{eqnarray}
where  $M=\frac{1}{2}+\frac{1}{2(2N+1)}[(1-q)^{(2N+1)}-1]$, $\lambda_{\pm}=\frac{1}{2}\pm\sqrt{\frac{1}{4}(1-q)^{(2N+1)}+(M-\frac{1}{2})^{2}}$, and $q=1-e^{-\gamma_0\tau}$ donates the noisy parameter.

\begin{figure}[htbp]
\centering
\includegraphics[height=2.0in,width=1.6in]{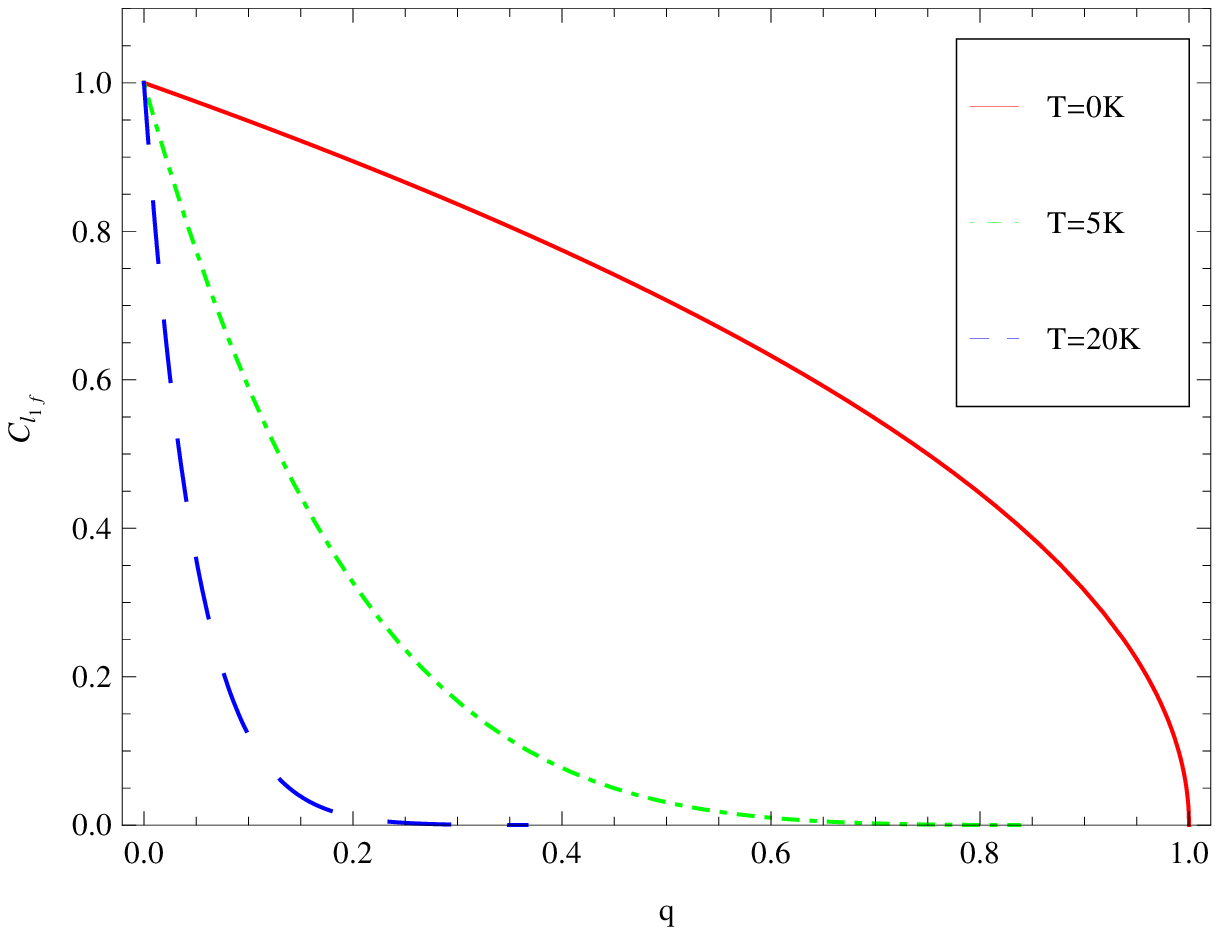}
\includegraphics[height=2.0in,width=1.6in]{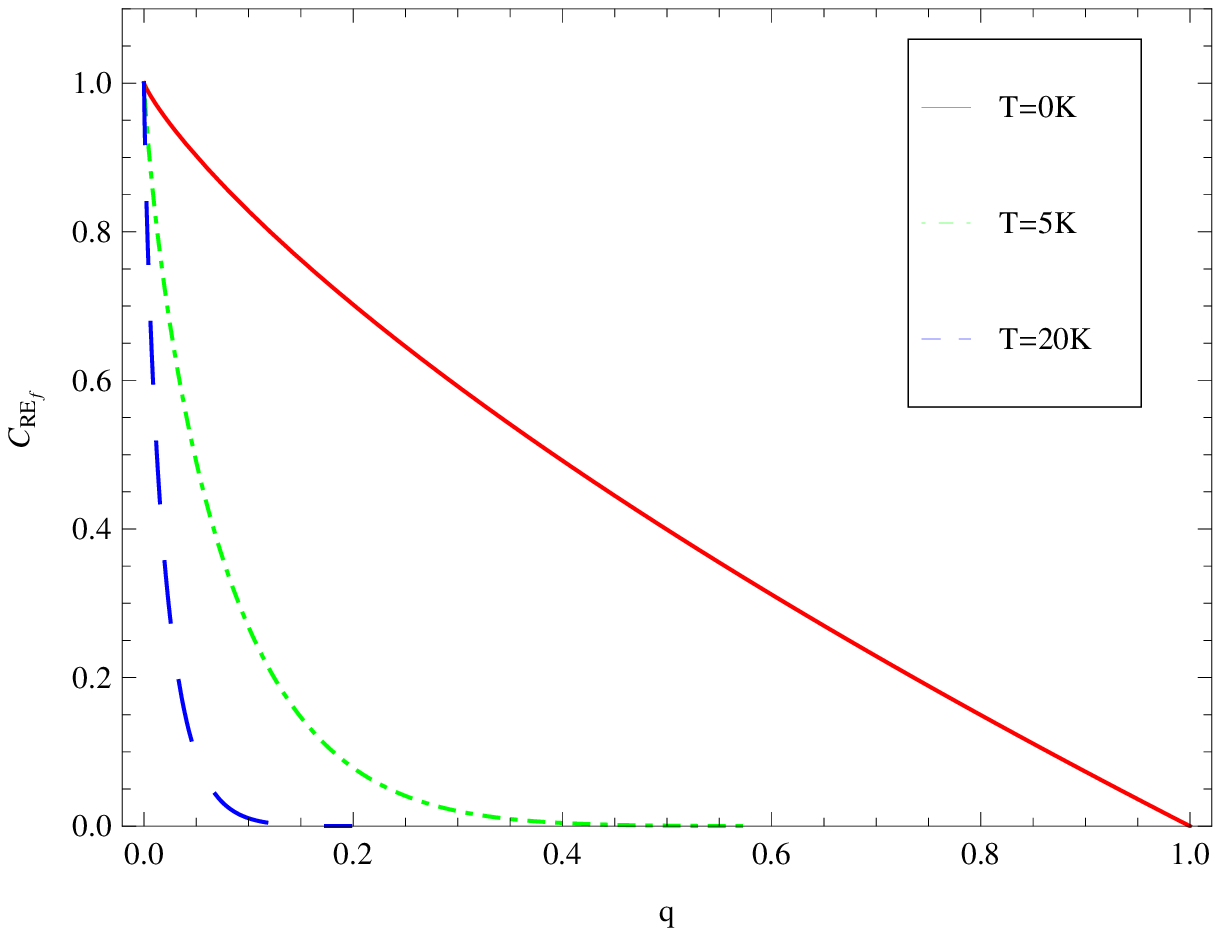}
\caption{ (color online). The \textit{$l_1$} norm and relative entropy of coherence respectively are shown as the function of $q$ at different temperature $T$. Here, we take fixed $T = 0K,\; 5K,\; 20K$ respectively.
}\label{f1}
\end{figure}

To show how the noisy environments affect the quantum coherence,  we plot the \textit{$l_1$} norm and relative entropy of coherence respectively as a function of $q$ at different temperature $T$  in Fig.~(\ref{f1}). We can see that the quantum coherence decreases monotonically as the evolution time goes by due to the decoherence caused by the unavoidable interaction between the atom and the electromagnetic field in a thermal bath with temperature $T$.
What is important is that the quantum coherence decreases faster as the temperature $T$ increases. Compared with the behavior of the coherence of the atom coupled with the fluctuating vacuum field~\cite{Liu}, we find that $\tau_f$  decreases as the temperature  increases, which means that the quantum coherence at higher temperature decays more rapidly than those for zero temperature.

\section{inhibiting the decoherence by the presence of a reflecting boundary}
As shown above, the atom interacting with a thermal bath of electromagnetic field will be inevitably subjected to decoherence even though the field is in the Minkowski vacuum. Thus it is needed to analyze how to protect quantum coherence in this setting. To solve this problem, we will put a reflecting boundary in the free space and discuss the dynamics of the quantum coherence of the two-level atom again. In this case, we assume the boundary is located at $z=0$, then the electric field two-point function in the Feynman gauge~\cite{Brown,Birrell} can be write as
\begin{widetext}
\begin{eqnarray}\label{boundary}
\langle\textbf{E}_{i}(x(\tau))\textbf{E}_{j}(x(\tau'))\rangle&=&\frac{\hbar c}{4\pi^{2}\varepsilon_{0}}\sum_{n=-\infty}^{\infty}\frac{\delta_{ij}\partial_{0}\partial_{0}'-\partial_{i}\partial_{j}'}{(x-x')^{2}+(y-y')^{2}+(z-z')^{2}-(c t-c t'+i n\beta-i\varepsilon)^{2}}\nonumber\\
&-&\frac{\hbar c}{4\pi^{2}\varepsilon_{0}}\sum_{n=-\infty}^{\infty}\frac{[(\delta_{ij}-2n_{i}n_{j})\partial_{0}\partial_{0}'-\partial_{i}\partial_{j}']}{(x-x')^{2}+(y-y')^{2}+(z+z')^{2}-(c t-c t'+i n\beta-i\varepsilon)^{2}}\;,
\end{eqnarray} \end{widetext}
which is a sum of the unbounded space term and the boundary-dependent term. Here, $\textbf{n}=(0,0,1)$ donates the unit vector normal to the reflecting boundary. For the rest atom, submitting the trajectory in Eqs.~(\ref{trajectory}) and  (\ref{boundary})  into Eq.~(\ref{transforms}), we can calculate the Fourier transform  of the two-point function of the electromagnetic field at finite temperature in the presence of a boundary as follows\begin{widetext}
\begin{eqnarray}\label{Fourier2}
\mathcal{G}_{b}(\lambda)=
\begin{cases}
\;\sum_{i=1}^{3}\frac{e^{2}|\langle-|r_{i}|+\rangle^{2}\lambda^{3}}{3\pi\varepsilon_{0}\hbar c^{3}}[1-f_{i}(\lambda,z)](1+N(\lambda)),\;\;(\lambda>0)\;,\\
\;\sum_{i=1}^{3}\frac{e^{2}|\langle-|r_{i}|+\rangle^{2}|\lambda|^{3}}{3\pi\varepsilon_{0}\hbar c^{3}}[1-f_{i}(\lambda,z)]N(|\lambda|)),\;\;(\lambda <0)\;,
\end{cases}
\end{eqnarray}\end{widetext}
where the subscript `` b " donates the part induced by the presence of a reflecting boundary and we have defined
\begin{eqnarray}
f_x(\lambda,z_0)&=&f_y(\lambda,z_0)\nonumber \\
&=&\frac{3c^3}{16\lambda^3z_0^3}\bigg[\frac{2\lambda z_0}{c}\cos\frac{2\lambda z_0}{c}+\bigg(\frac{4\lambda^2z_0^2}{c^2}-1\bigg)\sin\frac{2\lambda z_0}{c}\bigg]\;,\nonumber\\
f_z(\lambda,z_0)&=&\frac{3c^3}{8\lambda^3z_0^3}\bigg[\frac{2\lambda z_0}{c}\cos\frac{2\lambda z_0}{c}-\sin\frac{2\lambda z_0}{c}\bigg]\;,
\end{eqnarray}
where $z_0$ is the atomic position from the boundary and we take the unit of $c/\omega_0$ in the figures.
The coefficients of the Kossakowski matrix $a_{ij}$ in the presence of a reflecting boundary are then given by
\begin{eqnarray}
A_b&=&\frac{\gamma_0}{4}\Big[1-\sum_i\;\alpha_if_i(\omega_0,z_0)\Big](2N+1)\;,\nonumber\\
B_b&=&\frac{\gamma_0}{4}\Big[1-\sum_i\;\alpha_if_i(\omega_0,z_0)\Big]\;,
\end{eqnarray}
where $\alpha_i=|\langle-|r_i|+\rangle|^2/|\langle -|{\bf r}|+\rangle|^2$ is defined as the atomic relative polarizability and satisfies $\sum_i\alpha_i=1$, and $\gamma_{b}=\gamma_0(2N+1)[1-\sum_i\;\alpha_if_i(\omega_0,z_0)\big]$ is the total transition rate in the presence of a boundary.  We let $f(\omega_0,z_0)\equiv\sum_i\alpha_if_i(\omega_0,z_0)$ hereafter for the sake of convenience.  When the temperature is zero, the emission rate in the presence of a boundary becomes $\gamma_0[1-f(\omega_0,z_0)]$ which is in agreement with the expression in Ref.~\cite{Meschede}. Let us note that here $\tau_b=\frac{1}{2\gamma_0(2N+1)[1-f(\omega_0,z_0)]}$ describes the time scale for the off-diagonal elements of the density-matrix decay in the presence of a boundary. Then, similar to the analysis before,  the \textit{$l_1$} norm and the relative entropy of coherence for the two-level atom by the presence of a reflecting boundary can be calculated as
\begin{eqnarray}\label{norm1}
C_{l_1}(\hat{\rho})_b=\left|e^{-\frac{\tau}{\tau_{b}}} \right|=|(1-q)^{(N+\frac{1}{2}) [1-f(\omega_0,z_0)]}|\;,
\end{eqnarray}
and
\begin{eqnarray}\label{entropy1}
\nonumber
C_{\rm RE}(\hat{\rho})_b&=&-L\log_{2}L-(1-L)\log_{2}(1-L)
\\
&&+\kappa_{+}\log_{2}\kappa_{+}+\kappa_{-}\log_{2}\kappa_{-}\;,
\end{eqnarray}
where $L=\frac{1}{2}+\frac{1}{2(2N+1)}[(1-q)^{(2N+1)[1-f(\omega_0,z_0)]}-1]$ and $\kappa_{\pm}=\frac{1}{2}\pm\sqrt{\frac{1}{4}(1-q)^{(2N+1) [1-f(\omega_0,z_0)]}+(L-\frac{1}{2})^{2}}$.

\begin{figure}[htbp]
\centering
\includegraphics[height=1.8in,width=3.4in]{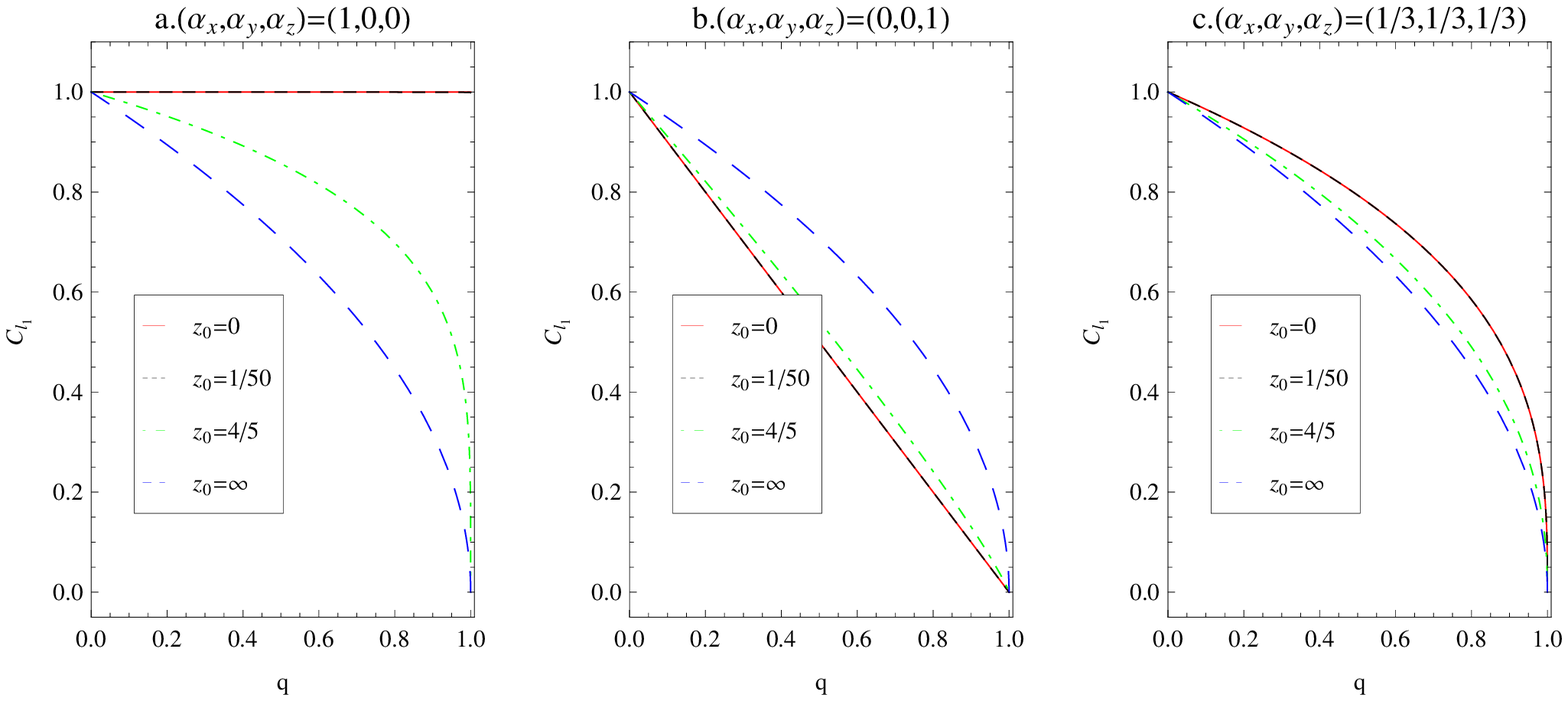}
\caption{ (color online). The \textit{$l_1$} norm of coherence in the presence of a reflecting boundary is shown as the function of $q$ at temperature $T=0 K$ when the atom is at different position $z_{0}$ for $(\alpha_x, \alpha_y, \alpha_z)=(1,0,0),\;(0,0,1),\;(1/3,1/3,1/3)$, which is shown in $(a),\;(b),\;(c)$ respectively. Here, we take fixed $z_{0} = 0,\;1/50,\; 4/5,\; \infty$ respectively and the values of all the curves are in the unit of $c/\omega_0$.
}\label{f2}
\end{figure}

\begin{figure}[htbp]
\centering
\includegraphics[height=1.8in,width=3.4in]{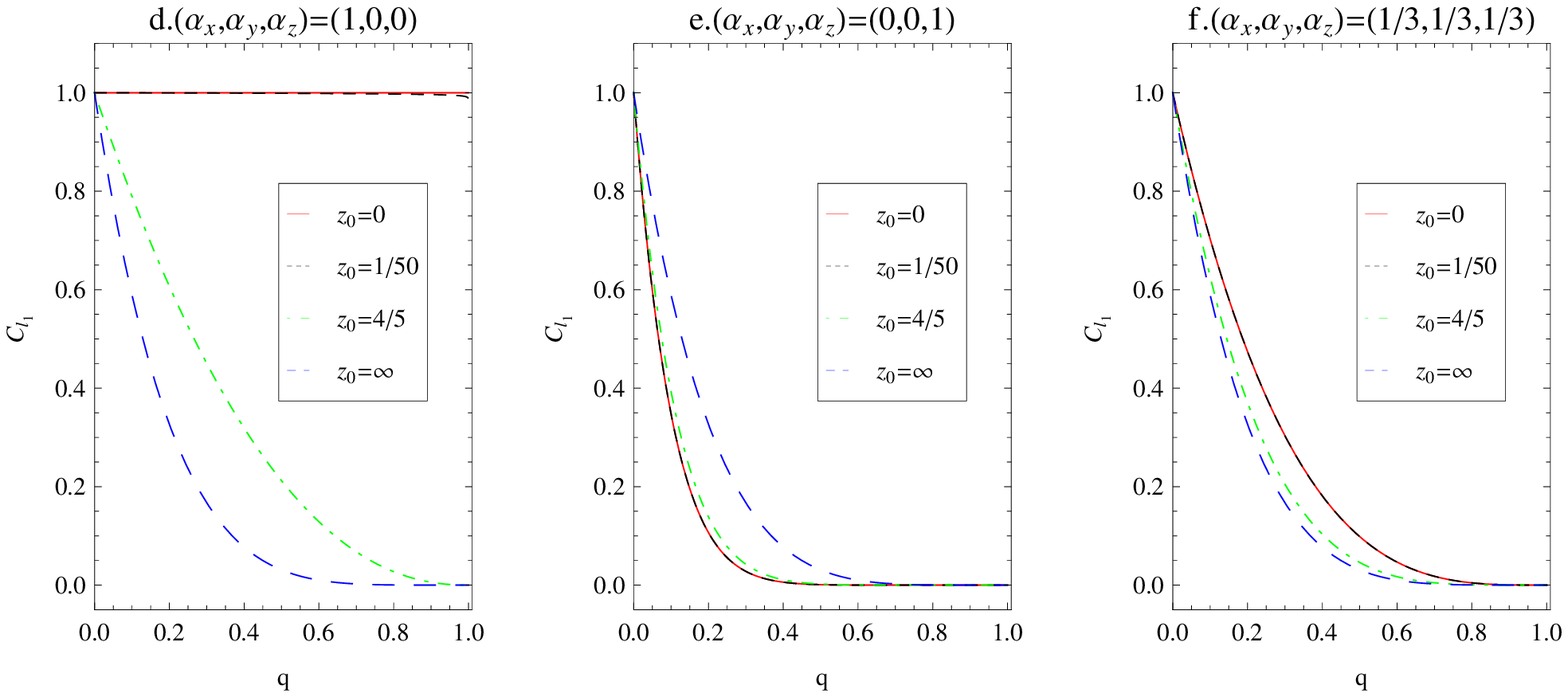}
\caption{ (color online). The \textit{$l_1$} norm of coherence in the presence of a reflecting boundary is shown as the function of $q$ at temperature $T=5 K$ when the atom is at different position $z_{0}$ for $(\alpha_x, \alpha_y, \alpha_z)=(1,0,0),\;(0,0,1),\;(1/3,1/3,1/3)$, which is shown in $(d),\;(e),\;(f)$  respectively. Here, we take fixed $z_{0} = 0,\;1/50,\; 4/5,\; \infty$ respectively and the values of all the curves are in the unit of $c/\omega_0$.
}\label{f3}
\end{figure}

\begin{figure}[htbp]
\centering
\includegraphics[height=1.8in,width=3.4in]{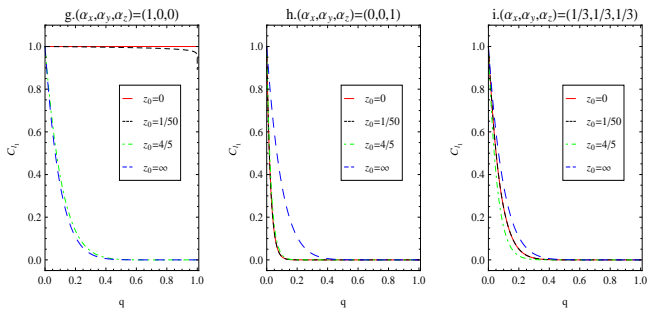}
\caption{ (color online). The \textit{$l_1$} norm of coherence in the presence of a reflecting boundary is shown as the function of $q$ at temperature $T=20K$ when the atom is at different position $z_{0}$ for $(\alpha_x, \alpha_y, \alpha_z)=(1,0,0),\;(0,0,1),\;(1/3,1/3,1/3)$, which is  shown in $(g),\;(h),\;(i)$  respectively. Here, we take fixed $z_{0} = 0,\;1/50,\; 4/5,\; \infty$ respectively and the values of all the curves are in the unit of $c/\omega_0$.
}\label{f4}
\end{figure}

In the following we will focus our attention on the conditions under which the \textit{$l_1$} norm of coherence is protected effectively in the presence of a reflecting boundary, which means that the quantum coherence can be protected as the evolution time  increases. Comparing with the unbound case, we can see that the \textit{$l_1$} norm of coherence in Eq.~(\ref{norm1}) is modified by the factor $1-f(\omega_{0},z_{0})$.
In order to study the properties of the \textit{$l_1$} norm of coherence $C_{l_1}(\hat{\rho})_b$ at different temperature $T$, we consider that the atom is located at different position with parallel, vertical, isotropic polarization, respectively. The behavior of the $C_{l_1}(\hat{\rho})_b$ in the presence of a reflecting boundary are shown in Figs.~(\ref{f2} - \ref{f4}), which is the function of $q$ at different temperature $T$ for $(\alpha_x, \alpha_y, \alpha_z)=(1,0,0),\;(0,0,1),\;(1/3,1/3,1/3)$ respectively and the atom is placed at different position $z_{0}$.


When the atom is far away from the reflecting boundary, i.e., $z_{0}\rightarrow\infty$, the term $1-f(\omega_{0},z_{0})\rightarrow 1$ in expressions (\ref{norm1}) and (\ref{entropy1}), and we immediately recover the atom at rest in the unbounded Minkowski case. For the case of the atom is very close to the reflecting boundary ( $z_{0}\rightarrow 0$), and the atom has parallel dipole polarization, we obtain $\alpha_{z}=0$ and $f_x(\omega_0,z_0)=f_y(\omega_0,z_0)=1$. This results in the total transition rate $\gamma_b$ becomes zero and  $\tau_b\rightarrow \infty$  for any finite temperature $T$, which means that  the decoherence is inhibited effectively by the presence of a boundary. In other words, the resource of  the quantum coherence is  protected effectively at any finite temperature $T$ under these conditions. When the atom is close to the boundary and has normal dipole polarization, i.e., $\alpha_{x}=\alpha_{y}=0$ and $f_z(\omega_0,z_0)=-1$, we find the total transition rate $\gamma_b$ is double of that the unbounded case and $\tau_b=\tau_{f}/2$,  which means in this case, the quantum coherence will be destroy faster than that without boundaries. However, at the same temperature $T$,  the decoherence decreases with the increase of the distance of the atom from the boundary, which makes the quantum coherence get lost slowly. When the atom is close to the boundary and has an isotropic dipole polarization, i.e., $\alpha_{x}=\alpha_{y}=\alpha_{z}=1/3$ and $f_x(\omega_0,z_0)=f_y(\omega_0,z_0)=-f_z(\omega_0,z_0)=1$, we have the total transition rate $\gamma_b=2\gamma_{f}/3$ and  $\tau_b=3\tau_{f}/2$. That is to say,   the decoherence  is inhibited to a certain degree and the quantum coherence decreases slowly  by the presence of the reflecting boundary  as compared to the unbounded case.

Noting that  for a different atomic position and polarization at different temperature, the decoherence may be inhibited, enhanced or unchanged as compared to the unbounded case. In above discussions, we find there exist a special position which has $1-f(\omega_{0},z_{0})=1$, making the \textit{$l_1$} norm of coherence is  similar to that in the unbounded space at these position, which means the boundary effects vanish. In addition, we can find a special position where $1-f(\omega_{0},z_{0})=0$, so the quantum coherence will be protected effectively.  While the atom is  placed at different position between the two special positions, the decoherence is inhibited or enhanced as compared to that the unbounded case, i.e., $1-f(\omega_{0},z_{0})> 1$ or $1-f(\omega_{0},z_{0})< 1$.  More importantly, we can see from the Figs.~(\ref{f2} - \ref{f4}) that  the \textit{$l_1$} norm of coherence is damaged more seriously  as the temperature increases in the same situation.
It is worthy to notice that the behaviors of  the relative entropy of coherence are similar to that  the \textit{$l_1$} norm of coherence, so we won't discuss them in detail.

\section{inhibiting the decoherence by the presence of two reflecting boundaries}
We have shown that under certain conditions, quantum coherence of a two-level atom can be effectively protected from the thermal bath if there is a boundary. Thus, it is interesting to ask what happens to the quantum coherence of two-level atom when two reflecting boundaries exist. In this case, we assume that these two boundaries are parallel with each other and both of them are parallel to the $x-y$ plane.  One is at $z=0$, and the other is at $z=a$. The two-level atom is located between them with the trajectory (\ref{trajectory}) and $z_0\in[0, a]$. In the presence of two perfectly reflecting boundaries, the electric field two-point function at finite temperature can be obtained by the method of images~\cite{Brown},
\begin{widetext}
\begin{eqnarray}\label{boundaries}
\langle\textbf{E}_{i}(x(\tau))\textbf{E}_{j}(x(\tau'))\rangle&=&\frac{\hbar c}{4\pi^{2}\varepsilon_{0}}\sum_{m=-\infty}^{\infty}\sum_{n=-\infty}^{\infty}\frac{\delta_{ij}\partial_{0}\partial_{0}'-\partial_{i}\partial_{j}'}{(x-x')^{2}+(y-y')^{2}+(z-z'-2 a m)^{2}-(c t-c t'+i n\beta-i\varepsilon)^{2}}\nonumber\\
&-&\frac{\hbar c}{4\pi^{2}\varepsilon_{0}}\sum_{m=-\infty}^{\infty}\sum_{n=-\infty}^{\infty}\frac{[(\delta_{ij}-2n_{i}n_{j})\partial_{0}\partial_{0}'-\partial_{i}\partial_{j}']}{(x-x')^{2}+(y-y')^{2}+(z+z'-2 a m)^{2}-(c t-c t'+i n\beta-i\varepsilon)^{2}}\;,
\end{eqnarray}\end{widetext}
which consists of a sum of the free space and a term that depends on the presence of the boundaries. Substituting the Eqs.~(\ref{trajectory})  and  (\ref{boundaries}) into  Eq.~(\ref{transforms}), we can obtain the Fourier transform of the electromagnetic field correlation function in the frame of the static atom, which is given by\begin{widetext}
\begin{eqnarray}\label{Fourier3}
\mathcal{G}_{b}(\lambda)=
\begin{cases}
\;\sum_{m=-\infty}^{\infty}[G(\lambda,am)-H(\lambda,z_0-am)][1+N(\lambda)],\;\;(\lambda>0)\;,\\
\;\sum_{m=-\infty}^{\infty}[G(\lambda,am)-H(\lambda,z_0-am)]N(|\lambda|),\;\;(\lambda <0)\;,
\end{cases}
\end{eqnarray}
\end{widetext}
where
\begin{eqnarray}
\nonumber
G(\lambda,am)&=&\bigg[\frac{e^{2}|\langle-|r_{x}|+\rangle^{2}\lambda^{3}}{3\pi\varepsilon_{0}\hbar c^{3}}f_{x}(\lambda,am)+\frac{e^{2}|\langle-|r_{y}|+\rangle^{2}\lambda^{3}}{3\pi\varepsilon_{0}\hbar c^{3}}
\\ \nonumber
&&\times\,f_{y}(\lambda,am)-\frac{e^{2}|\langle-|r_{z}|+\rangle^{2}\lambda^{3}}{3\pi\varepsilon_{0}\hbar c^{3}}f_{z}(\lambda,am)\bigg]\;,
\\ \nonumber
H(\lambda,z_0-am)&=&\sum_{i}\frac{e^{2}|\langle-|r_{i}|+\rangle^{2}\lambda^{3}}{3\pi\varepsilon_{0}\hbar c^{3}}f_{i}(\lambda,z_0-am)\;.
\end{eqnarray}
Here, the subscript `` t " represents the part induced by the presence of two reflecting boundaries. Consequently,
from the Fourier transform we can obtain the coefficients of the Kossakowski matrix $a_{ij}$ in the presence of two reflecting boundaries
\begin{eqnarray}
A_t&=&\frac{\gamma_0}{4}\sum_{m=-\infty}^{\infty}\big[G(\omega_0,am)-H(\omega_0,z_0-am)\big](2N+1)\;,\nonumber\\
B_t&=&\frac{\gamma_0}{4}\sum_{m=-\infty}^{\infty}\big[G(\omega_0,am)-H(\omega_0,z_0-am)\big]\;,
\end{eqnarray}
where $G(\omega_0,am)=\alpha_{x}f_{x}(\omega_0,am)+\alpha_{y}f_{y}(\omega_0,am)-
\alpha_{z}f_{z}(\omega_0,am)$ and $H(\omega_0,z_0-am)=\sum_{i}\alpha_{i}f_{i}(\omega_0,z_0-am)$. Here, $f_i$ and $\alpha_i$ with $i=x,y,z$ are given before.  Thus, the total transition rate of the two-level atom is $\gamma_t=\gamma_0\sum_{m=-\infty}^{\infty}\big[G(\omega_0,am)-H(\omega_0,z_0-am)\big](2N+1)$.  At zero temperature,  the emission rate $\gamma_0\sum_{m=-\infty}^{\infty}\big[G(\omega_0,am)-H(\omega_0,z_0-am)\big]$ which is obtained in this way is agree with the previous results~{\cite{Martini}}. It is noteworthy that $\tau_t=\frac{1}{2\gamma_0\sum_{m=-\infty}^{\infty}\big[G(\omega_0,am)-H(\omega_0,z_0-am)\big](2N+1)}$ is the time scale for the off-diagonal elements of the density-matrix decay in the presence of two reflecting boundaries. We can  obtain the \textit{$l_1$} norm and the relative entropy of coherence  in the presence of two reflecting boundaries
\begin{eqnarray}\label{norm2}
C_{l_1}(\hat{\rho})_t&=&|e^{-\frac{\tau}{\tau_t}}|=|(1-q)^{ (N+\frac{1}{2}) \sum^{\infty}_{m=-\infty}[G(\omega_0,am)-H(\omega_0,z_0-am)]}|\;,\nonumber\\
\end{eqnarray}
and
\begin{eqnarray}\label{entropy2}
C_{\rm RE}(\hat{\rho})_t&=&1+S\log_{2}S+(1-S)\log_{2}(1-S)\nonumber\\ &&-
\nu_{+}\log_{2}\nu_{+}-\nu_{-}\log_{2}\nu_{-}\;,
\end{eqnarray}
where
\begin{eqnarray}
S&=&\frac{1}{2}+\frac{(1-q)^{(2N+1)\sum^{\infty}_{m =-\infty}[G(\omega_0,am)-H(\omega_0,z_0-am)]}-1}{2(2N+1)}, \nonumber \\ \nu_{\pm}&=&\frac{1\pm\sqrt{(1-q)^{(2N+1) \sum^{\infty}_{m=-\infty}[G(\omega_0,am)-H(\omega_0,z_0-am)]} +(2S-1)^{2}}}{2}.\nonumber
\end{eqnarray}
\begin{figure}[htbp]
\centering
\includegraphics[height=2.0in,width=3.40in]{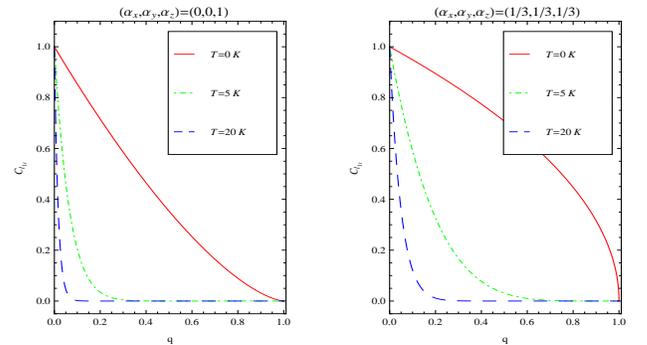}
\caption{ (color online). The \textit{$l_1$} norm of coherence is shown as the function of $q$ at different temperature $T$ when the atom is at arbitrary position $z_{0}$  for $(\alpha_x, \alpha_y, \alpha_z)=(0,0,1),\;(1/3,1/3,1/3)$ respectively. Here, we take fixed $T = 0K,\; 5K, \;20K$, respectively and the values of all the curves are in the unit of $c/\omega_0$.
}\label{f5}
\end{figure}

Considering the \textit{$l_1$} norm  of coherence firstly, the term $\sum^{\infty}_{m=-\infty}[G(\omega_0,am)-H(\omega_0,z_0-am)]$
in the Eq.~(\ref{norm2}) will modify the quantum coherence  by the presence of boundaries. The  \textit{$l_1$} norm  of coherence not only depends on the atomic position $z_0$ and  polarization, but also  depends on the temperature and the distance between two boundaries. In this section we only consider three situations of the distance between the two boundaries, which takes fixed $a=\pi/2,\;3\pi/2,\;\infty$ ( we used unit of $c/\omega_0$).

\emph{ (i)} We start with the limiting when the distance between the two boundaries is infinite, i.e., $a\rightarrow\infty$. In this case, we recover the electromagnetic field at finite temperature $T$ in the presence of a reflecting boundary as expected which we have discussed in Sec. IV.

\emph{ (ii)} For the case when the distance between the two boundaries $a=\pi/2$, it is deserved our attention that the \textit{$l_1$} norm  of coherence has nothing to do with the atomic position $z_0$ between the two parallel boundaries.
Because of  $\sum_{m=-\infty}^{\infty}G(\omega_0,am)=\frac{3}{4}\alpha_x+\frac{3}{4}\alpha_y+\frac{3}{2}\alpha_z$ and  $\sum_{m=-\infty}^{\infty}H(\omega_0,z_0-am)=\frac{3}{4}\alpha_x+\frac{3}{4}\alpha_y-\frac{3}{2}\alpha_z$ in Eq.~(\ref{norm2}),  the $l_1$ norm of quantum coherence can be obtained
\begin{eqnarray}
C_{l_1}(\hat{\rho})_t=|(1-q)^{ (N+\frac{1}{2}) [(\frac{3}{4}\alpha_x+\frac{3}{4}\alpha_y+\frac{3}{2}\alpha_z)-(\frac{3}{4}\alpha_x+\frac{3}{4}\alpha_y-\frac{3}{2}\alpha_z)]}|\;,
\end{eqnarray}
from which we know that the $C_{l_1}(\hat{\rho})_t$ only depends on the atomic polarization and the temperature $T$. When the dipole moment is parallel to the $xy$ plane, i.e., $\alpha_z=0$, the total transition rate $\gamma_t=0$ and  $\tau_t\rightarrow\infty$ at any finite temperature,  which means that the  decoherence is totally inhibited by the presence of two reflecting boundaries and the important quantum resources of coherence is  protected totally  wherever the atom is placed between the boundaries. For a dipole moment perpendicular to boundaries, i.e., $\alpha_x= \alpha_y=0$, we have the total transition rate $\gamma_t=3\gamma_f$ and $\tau_t=\tau_f/3$, which implies that the quantum coherence damages faster  than that in the unbounded case no matter where the atom is placed.  When the dipole moment has an isotropic polarization  $\alpha_x= \alpha_y=\alpha_z=1/3$, we can get the total transition rate $\gamma_{t}=\gamma_{f}$ and $\tau_t=\tau_f$, which means that the effects of the boundaries has been eliminated and the unbounded case is recovered.

The decoherence may be enhanced, inhibited or even not changed by the presence of two reflecting boundaries as compared to the unbounded case, which  only depends on  the atomic polarization and the temperature $T$. It is interesting to find that the decoherence is inhibited totally as long as the polarization is in the $xy$ plane at any finite temperature.  Then, in Fig.~(\ref{f5}) we plot the \textit{$l_1$} norm  of coherence $C_{l_1}(\hat{\rho})_t$, in the presence of two reflecting boundaries, as a function of $q$ at different temperature $T$ for $(\alpha_x, \alpha_y,\alpha_z)=(0,0,1),\;(1/3,1/3,1/3)$  respectively.
It is easy to find that the \textit{$l_1$} norm  of coherence is damaged faster with the increase of temperature $T$.

\emph{ (iii)} Finally, for the case when the distance between the two boundaries $a=3\pi/2$, we find that the decoherence is enhanced or inhibited as compared to the unbounded case, which is related to not only the atomic  position $z_0$ but also the atomic polarization direction and the temperature. Interestingly, we find the properties of the \textit{$l_1$} norm  of coherence  are symmetrical about $z=a/2$. Therefore, we  analyze the range from $z_0=0$ to $z_0=a/2$ in the following.

\begin{figure}[htbp]
\centering
\includegraphics[height=1.8in,width=3.4in]{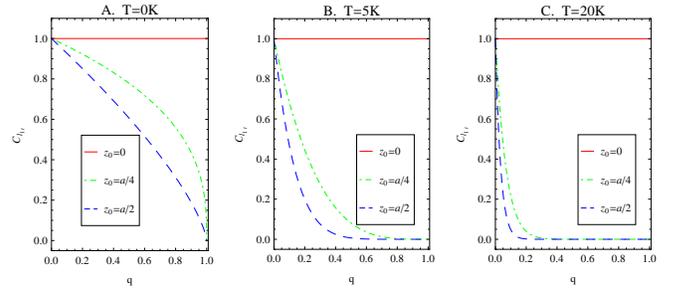}
\caption{ (color online). The  \textit{$l_1$} norm of coherence in the presence of two reflecting boundaries is shown as the function of $q$ with temperature $T=0K,\;5K,\;20K$  for $(\alpha_x, \alpha_y, \alpha_z)=(1,0,0)$  at different atomic position $z_{0}$, which is  shown in $(A),\;(B),\;(C)$  respectively.  Here, we take fixed $z_{0} = 0,\; a/4,\; a/2$ respectively and the values of all the curves are in the unit of $c/\omega_0$.
}\label{f6}
\end{figure}

\emph{(a)} For a dipole oriented parallel to the boundary, the relation $\alpha_z=0$ is satisfied.  When the atom is placed very close to the boundary ($z_0\rightarrow0$), we have the total transition rate $\gamma_t=0$ and $\tau_t \rightarrow \infty$, which represents the decoherence can be inhibited effectively and the quantum coherence will be protected effectively  at any finite temperature. When the atom is placed at $z_0=a/4$, the total transition rate $\gamma_t$ is $13\gamma_f/18$ and $\tau_t=18\tau_f/13$, which means that the decoherence can be inhibited to some degree as compared to the unbounded case and the quantum coherence is destroyed more slowly. When the atom is placed at $z_0=a/2$, we have $\gamma_t=13\gamma_f/9$ and $\tau_t=9\tau_f/13$, leading to the decoherence enhanced  and the quantum coherence damaged faster  than that in unbounded case. As a result,  in the ranges of  $0\leq z_0\leq a/2$, we can find that the total transition rate $\gamma_t$  increases but the $\tau_t$ decreases, resulting in the  decoherence is  enhanced with the increase of $z_0$. To show the properties of the \textit{$l_1$} norm  of coherence, we plot it in Fig.~(\ref{f6}) as a function of $q$  at different atomic position $z_0$ for $(\alpha_x, \alpha_y, \alpha_z)=(1,0,0)$ with temperature $T=0K,\;5K,$ and $20K$, respectively. The quantum coherence displayed in Fig.~(\ref{f6}) is damaged faster  with the temperature increasing.
\begin{figure}[htbp]
\centering
\includegraphics[height=1.8in,width=3.4in]{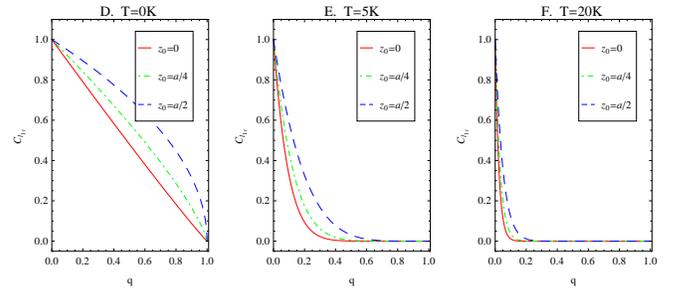}
\caption{ (color online). The  \textit{$l_1$} norm of coherence in the presence of two reflecting boundaries is shown as the function of $q$ with temperature $T=0K,\;5K,\;20K$ for  $(\alpha_x, \alpha_y, \alpha_z)=(0,0,1)$ at different atomic position $z_{0}$, which is  shown in $(D),\;(E),\;(F)$  respectively.  Here, we take fixed $z_{0} = 0,\; a/4,\; a/2$ respectively and the values of all the curves are in the unit of $c/\omega_0$.
}\label{f7}
\end{figure}

\emph{(b)} For a dipole oriented normal to the boundary, i.e., $\alpha_x= \alpha_y=0$. When the atom is placed very close to the boundary ($z_0\rightarrow0$), the total transition rate $\gamma_t=19\gamma_f/9$ and the relaxation time $\tau_t=9\tau_f/19$, so the decoherence is enhanced  and the quantum coherence is greatly damaged as compared to the unbounded case. When the atom is placed at $z_0=a/4$, we can get $\gamma_t=14\gamma_f/9$ and $\tau_t=9\tau_f/14$, which results in the quantum coherence is destroyed more than that in the unbounded case. However, when the atom in the $z_0=a/2$, we have $\gamma_t=\gamma_f$ and $\tau_t=\tau_f$, i.e., the unbounded case is recovered. In this case, we find that the total transition rate $\gamma_t$  decreases while the  $\tau_t$  increases in the ranges of  $0\leq z_0\leq a/2$, leading to the  decoherence decreases with the increase of $z_0$. We plot the \textit{$l_1$} norm  of coherence in Fig.~(\ref{f7}), which is the function of $q$ at different atomic position $z_0$ for $(\alpha_x, \alpha_y, \alpha_z)=(0,0,1)$ with temperature $T=0K,\;5K,$ and $20K$ respectively.  We  can see from the Fig.~(\ref{f7}) that the quantum coherence is also destroyed faster with the increase of temperature.
\begin{figure}[htbp]
\centering
\includegraphics[height=1.8in,width=3.4in]{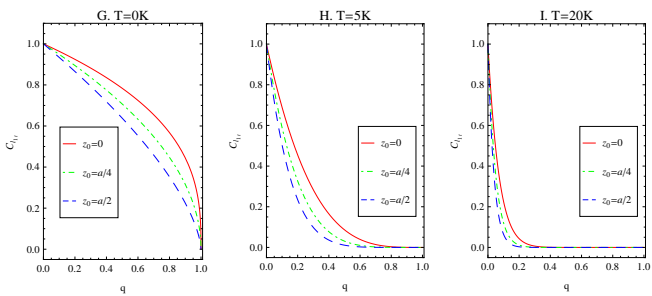}
\caption{ (color online). The  \textit{$l_1$} norm of coherence in the presence of two reflecting boundaries is shown as the function of $q$ with temperature $T=0K,\;5K,\;20K$ for  $(\alpha_x, \alpha_y, \alpha_z)=(1/3,1/3,1/3)$ at different atomic position $z_{0}$, which is  shown in $(G),\;(H),\;(I)$  respectively.  Here, we take fixed $z_{0} = 0, a/4, a/2$ respectively and the values of all the curves are in the unit of $c/\omega_0$.
}\label{f8}
\end{figure}

\emph{(c)} For a dipole has an isotropic polarization, i.e., $\alpha_x= \alpha_y=\alpha_z=1/3$. When the atom is placed very close to the boundary ($z_0\rightarrow0$), the total transition rate  $\gamma_t=19\gamma_f/27$ and $\tau_t=27\tau_f/19$, so the decoherence is inhibited to some degree as compared to the unbounded case. If the atom is placed at $z_0=a/4$, the total transition rate $\gamma_t=\gamma_f$ and  $\tau_t=\tau_f$, which leads to the bounded case reduces to the unbounded case as expected. If the atom is placed at $z_0=a/2$, we can get the total transition rate $\gamma_t=35\gamma_f/27$ and  $\tau_t=27\tau_f/35$, which means that the decoherence  enhances and the quantum coherence decreases faster than that the unbounded case. Thus, the $\tau_t$  decreases  as the total transition rate $\gamma_t$  increases, resulting in the decoherence  enhances  with the atomic position $z_0$ increases. The \textit{$l_1$} norm  of coherence is depicted in Fig.~(\ref{f8}) as a function of $q$ at different atomic position $z_0$ for $(\alpha_x, \alpha_y, \alpha_z)=(1/3,1/3,1/3)$ with temperature $T=0K,\;5K,$ and $20K$ respectively. Similarly, we can see from the Fig.~(\ref{f8}) that the quantum coherence decreases faster as the temperature increases. As expected, we noted that the properties of the relative entropy of coherence is similar to the \textit{$l_1$} norm  of coherence.

\section{Conclusions}
We have concerned about the conditions under which the decoherence can be inhibited effectively from noise environment in the open quantum systems. We have studied the dynamics of the quantum coherence of a two-level atom coupled with the electromagnetic field with the finite temperature. In this paper three cases are considered detaily: without boundaries, with the presence of a boundary, and the presence of two parallel boundaries.
In the case without boundaries, the  decoherence  can not be inhibited. Besides, the $\tau_f$  decreases when the temperature increases, i.e., the higher temperature leads to a faster decoherence.
In the case with the presence of a boundary, the quantum coherence  not only depends  on the temperature, but also depends on the atomic position and polarization direction. We found that the $\tau_b$ may be decreased, increased or even remain unchanged as compared to the unbounded case. That is to say, the decoherence may be enhanced, inhibited or unchanged as compared to the results for the free space. Only when the atom is close to the boundary and the transversely polarization, the $\tau_b\rightarrow\infty$ and the decoherence can be inhibited effectively at any finite temperature, which makes the quantum coherence can be protected highly.
And in the case with the presence of two parallel boundaries, we find that the quantum coherence is also related to  the distance between the two boundaries. Two different situations, the distance between the boundaries is $\pi/2$ or $3\pi/2$, are considered carefully. When the distance between the two boundaries is $\pi/2$, we  find that the quantum coherence has nothing to do with the atomic position  and the decoherence can be inhibited totally at any finite temperature when the atomic polarization is in the $xy$ plane, which leads to the quantum coherence is shielded from the influence of environment. However, when the distance between the boundaries is $3\pi/2$, the decoherence can be inhibited effectively only when the atom is close to the boundary and is transversely polarizable.

\begin{acknowledgments}
This work is supported by the  National Natural Science Foundation
of China under Grant Nos. 11475061 and 11305058;   the Open Project Program of State Key Laboratory of
Theoretical Physics, Institute of Theoretical Physics, Chinese
Academy of Sciences, China (No.Y5KF161CJ1).
\end{acknowledgments}


\begin{thebibliography}{00}
\bibitem{Leggett} A. J. Leggett, Prog. Theor. Phys. Suppl. {\bf69}, 80 (1980).
\bibitem{Glauber} R. J. Glauber, Phys. Rev. {\bf131}, 2766 (1963).
\bibitem{Sudarshan} E. C. G. Sudarshan, Phys. Rev. Lett. {\bf10}, 277 (1963).

\bibitem{Nielsen} M. Nielsen and I. Chuang, Quantum Computation and Quantum Information (Cambridge University Press, Cambridge, England, 2000), ISBN: 9781139495486.

\bibitem{Huelga} S. F. Huelga and M. B. Plenio, Contemp. Phys. {\bf54}, 181 (2013).
\bibitem{Marvian} I. Marvian and R. W. Spekkens, Phys. Rev. A. {\bf90}, 062110 (2014).
\bibitem{Bartlett} S. D. Bartlett, T. Rudolph, and R. W. Spekkens, Reviews of Modern Physics {\bf79}, 555 (2007).
\bibitem{Lostagio} M. Lostagio, K. Korzekwa, D. Jennings, and T.Rudolph, Phys. Rev. X. {\bf5}, 021001 (2015).
\bibitem{Lostaglio1} M. Lostaglio, . Jennings, and T. Rudolph, Nat. Commun. {\bf6}, 6383 (2015).
\bibitem{Giovannetti} V. Giovannetti, S. Lloyd, and L. Maccone, Science {\bf306}, 1330 (2004).
\bibitem{Demkowicz} R. Demkowicz-Dobrzanski and L. Maccone, Phys. Rev. Lett. {\bf113}, 250801 (2014).
\bibitem{berg} J. {\AA}berg, Phys. Rev. Lett. {\bf113}, 150402 (2014).
\bibitem{Baumgratz} T. Baumgratz, M. Cramer, and M. B. Plenio, Phys. Rev. Lett. {\bf113}, 140401 (2014).
\bibitem{Girolami} D. Girolami, Phys. Rev. Lett. {\bf113}, 170401 (2014).

\bibitem{Breuer} Heinz-Peter Breuer and Francesco Petruccione, The Theory of Open Quantum Systems (Oxford University, 2002).
\bibitem{Schlosshauer} M. Schlosshauer, Rev. Mod. Phys. {\bf76}, 1267 (2005).
\bibitem{Wang1} J. Wang and J. Jing, Phys. Rev. A {\bf82}, 032324 (2010).
\bibitem{Tian} Z. Tian and J. Jing, Phys. Lett. B {\bf707}, 264 (2012).
\bibitem{Liu} Xiaobao Liu, Zehua Tian, Jieci Wang, and Jiliang Jing, arXiv: 1509.06832.
\bibitem{Wang2} J. Wang, Z. Tian, J. Jing, and H. Fan, arXiv: 1601.03238.
\bibitem{Tian1} Z. Tian, J. Wang, H. Fan, and J. Jing, arXiv: 1501.06676.
\bibitem{Tian2} Z. Tian and J. Jing, Ann. Phys. {\bf350}, 1 (2014).
\bibitem{Jia} L. Jia, Z. Tian, and J. Jing, Ann. Phys. {\bf353}, 37 (2015).

\bibitem{thomas} Thomas R. Bromley, Marco Cianciaruso, and Gerardo Adesso, Phys. Rev. Lett. {\bf114}, 210401 (2015).
\bibitem{Cianciaruso} M. Cianciaruso, T. R. Bromley, W. Roga, R. Lo Franco, and G. Adesso, Sci. Rep. {\bf5}, 10177 (2015).
\bibitem{Silva} A. Silva et al., arxiv: 1511.01971.
\bibitem{Man} Z. Man, Y. Xia, and R. L. Franco, Scientific Reports {\bf5}, 13843 (2015).
\bibitem{Lo} R. Lo Franco, New J. Phys. {\bf17}, 081004 (2015).
\bibitem{Man1} Z. -X. Man, Y. -J. Xia, and R. Lo Franco, Phys. Rev. A {\bf92}, 012315 (2015).
\bibitem{Brito} F. Brito and T. Werlang, New. J. Phys. {\bf17}, 072001 (2015).
\bibitem{Rodriguez} F. J. Rodriguez, L. Quiroga, C. Tejedor, M. D. Martin, L. Vina, and R. Andre, Phys. Rev. B {\bf78}, 035312 (2008).
\bibitem{Gonzalez} A. Gonzalez-Tudela, F. J. Rodriguez, L. Quiroga, and C. Tejedor, Phys. Rev. B {\bf82}, 115334 (2010).
\bibitem{Guan} G. Wang, T. Li, and F. Deng, Quantum Inf. Process. {\bf14}, 1305 (2015).

\bibitem{Remote1} Z. Yang, T. Wu, J. Liu, 
 Acta Physica Sinica, (2016).
\bibitem{Remote2} X. Xiao, J. Xiao, Y. Ren, Y. Li, C. Ji, and X. Huang, 
Int. J. Theoret. Phys. 1 (2016).

\bibitem{Compagno} G. Compagno, R. Passante, and F. Persico, Atom-Field Interactions and Dressed Atoms (Cambridge University Press, Cambridge 1995).

\bibitem{Gorini} V. Gorini, A. Kossakowski, and E. C. G. Surdarshan, J. Math. Phys. {\bf17}, 821 (1976); G. Lindblad, Commun. Math. Phys. {\bf48}, 119 (1976).
\bibitem{Benatti} F. Benatti, R. Floreanini and M. Piani, Phys. Rev. Lett. {\bf91}, 070402 (2003).

\bibitem{Brown} L.S. Brown and G.J. Maclay, Phys. Rev. {\bf184}, 1272 (1969).
\bibitem{Yu} H. Yu, J. Chen, and P. Wu, J. High Energy Phys. {\bf02}, 058 (2006).

\bibitem{Birrell} N. D. Birrell, and P. C. W. Davies, Quantum Fields in Curved Space (Cambridge University, 1982).
\bibitem{Meschede} D. Meschede, W. Jhe and E. A. Hinds, Phys. Rev. A {\bf41}, 1587 (1990).

\bibitem{Martini} F. De Martini, M. Marrocco, P. Mataloni, L. Crescentini and R. Loudon, Phys. Rev. A {\bf43}, 2480 (1991).
\end{thebibliography}
\end{document}